\documentclass[twocolumn]{aastex63}
\pdfoutput=1
\usepackage{graphicx, url}
\usepackage{enumerate, enumitem}
\usepackage{amssymb, amsmath, amsfonts, upgreek}
\usepackage{gensymb}
\usepackage{mathtools}
\usepackage{hyperref}
\usepackage{xspace}
\usepackage{xcolor}
\usepackage{ulem}
\usepackage{booktabs}

\usepackage{savesym}
\savesymbol{tablenum}
\usepackage{siunitx}
\restoresymbol{SIX}{tablenum}

\usepackage{lineno}

\makeatletter
\newcommand\thefontsize[1]{{#1 The current font size is: \f@size pt\par}}
\makeatother

\newcommand{\reboundcodename}{{\tt REBOUND}\xspace}
\newcommand{\reboundxcodename}{{\tt REBOUNDx}\xspace}

\received{June 7, 2024}
\revised{Nov 1, 2024}
\accepted{Dec 4, 2024}

\shorttitle{HAT-P-11 Dynamical History}
\shortauthors{Lu et al.}
\submitjournal{ApJ}
\begin{document}

\title{Planet-Planet Scattering and ZLK Migration -- The Dynamical History of HAT-P-11}

\author[0000-0003-0834-8645]{Tiger Lu}
\affiliation{Department of Astronomy, Yale University, 219 Prospect Street, New Haven, CT 06511, USA}

\author[0000-0003-0115-547X]{Qier An}
\affiliation{Department of Physics, University of California, Santa Barbara, Santa Barbara, CA 93106, USA}
\affiliation{Department of Physics and Astronomy, Johns Hopkins University, Baltimore, MD 21218, USA}

\author[0000-0001-8308-0808]{Gongjie Li}
\affiliation{Center for Relativistic Astrophysics, School of Physics, Georgia Institute of Technology, Atlanta GA 30332, USA}

\author[0000-0003-3130-2282]{Sarah C. Millholland}
\affiliation{MIT Kavli Institute for Astrophysics and Space Research, Massachusetts Institute of Technology, Cambridge, MA 02139, USA}

\author[0000-0002-7670-670X]{Malena Rice}
\affiliation{Department of Astronomy, Yale University, 219 Prospect Street, New Haven, CT 06511, USA}

\author[0000-0003-0168-3010]{G.~Mirek Brandt}
\affiliation{Department of Physics, University of California, Santa Barbara, Santa Barbara, CA 93106, USA}

\author[0000-0003-2630-8073]{Timothy D.~Brandt}
\affiliation{Space Telescope Science Institute, 3700 San Martin Drive, Baltimore, MD 21218, USA}
\affiliation{Department of Physics, University of California, Santa Barbara, Santa Barbara, CA 93106, USA}

\begin{abstract}
The two planets of the HAT-P-11 system represent fascinating dynamical puzzles due to their significant eccentricities and orbital misalignments. In particular, HAT-P-11 b is on a close-in orbit that tides should have circularized well within the age of the system. Here we propose a two-step dynamical process that can reproduce all intriguing aspects of the system. We first invoke planet-planet scattering to generate significant eccentricities and mutual inclinations between the planets. We then propose that this misalignment initiated von-Zeipel-Lidov-Kozai cycles and high-eccentricity migration that ultimately brought HAT-P-11 b to its present-day orbit. We find that this scenario is fully consistent only when significant tidally-driven radius inflation is accounted for during the tidal migration. We present a suite of \textit{N}-body simulations exploring each phase of evolution and show that this scenario is consistent with all observational posteriors and the reported age of the system. 
\end{abstract}

\keywords{celestial mechanics, planet-star interactions, numerical methods}

\section{Introduction}
The mid-K dwarf HAT-P-11 hosts an intriguing system that has been the subject of much interest in recent years. The system has two known planets. The first, HAT-P-11 b, is a close-in ($a_\mathrm{b} = 0.0525$ AU) eccentric ($e_\mathrm{b} = 0.218$) super-Neptune ($m_\mathrm{b} = 23.4 \: \mathrm{M}_E$) first identified by \cite{Bakos+Torres+Pal+etal_2010}. Follow-up Rossiter-McLaughlin \citep{rossiter1924detection, mcLaughlin1924some} analysis of the system by \cite{hirano_2011} and \cite{2011ApJ_Sanchis-Ojeda_hatp11_star_spin} revealed that the orbit of HAT-P-11 b is polar ($\psi_\mathrm{Ab} = 106^\circ$), which makes it a member of the potential population of perpendicular planets \citep{albrecht_perponderance, Dong23, siegel_ponderings}. 

HAT-P-11 b presents an interesting dynamical puzzle. Planets as close-in as HAT-P-11 b are expected to be on circular orbits due to tidal forces. The fact that HAT-P-11 b is significantly eccentric, coupled with its unusual perpendicular orbit, points to a dynamically hot history. \cite{2018AJ_Yee_hatp11_RVS} greatly advanced our understanding of the dynamical history of the system with their discovery of the second planet in the system, HAT-P-11 c, an eccentric ($e_\mathrm{c} \approx 0.6$) super-Jupiter ($m_\mathrm{c} \sin i \approx 1.6 \: \mathrm{M}_J$) on an $a_\mathrm{c} \approx 4.1$ AU orbit. Their dynamical analysis of the system concluded that with this additional companion, both the eccentricity and spin-orbit misalignment of HAT-P-11 b could be explained on the condition that HAT-P-11 c was also misaligned. At the time, the orbital inclination of HAT-P-11 c was unconstrained.

\begin{figure}
    \centering
    \includegraphics[width=0.5\textwidth]{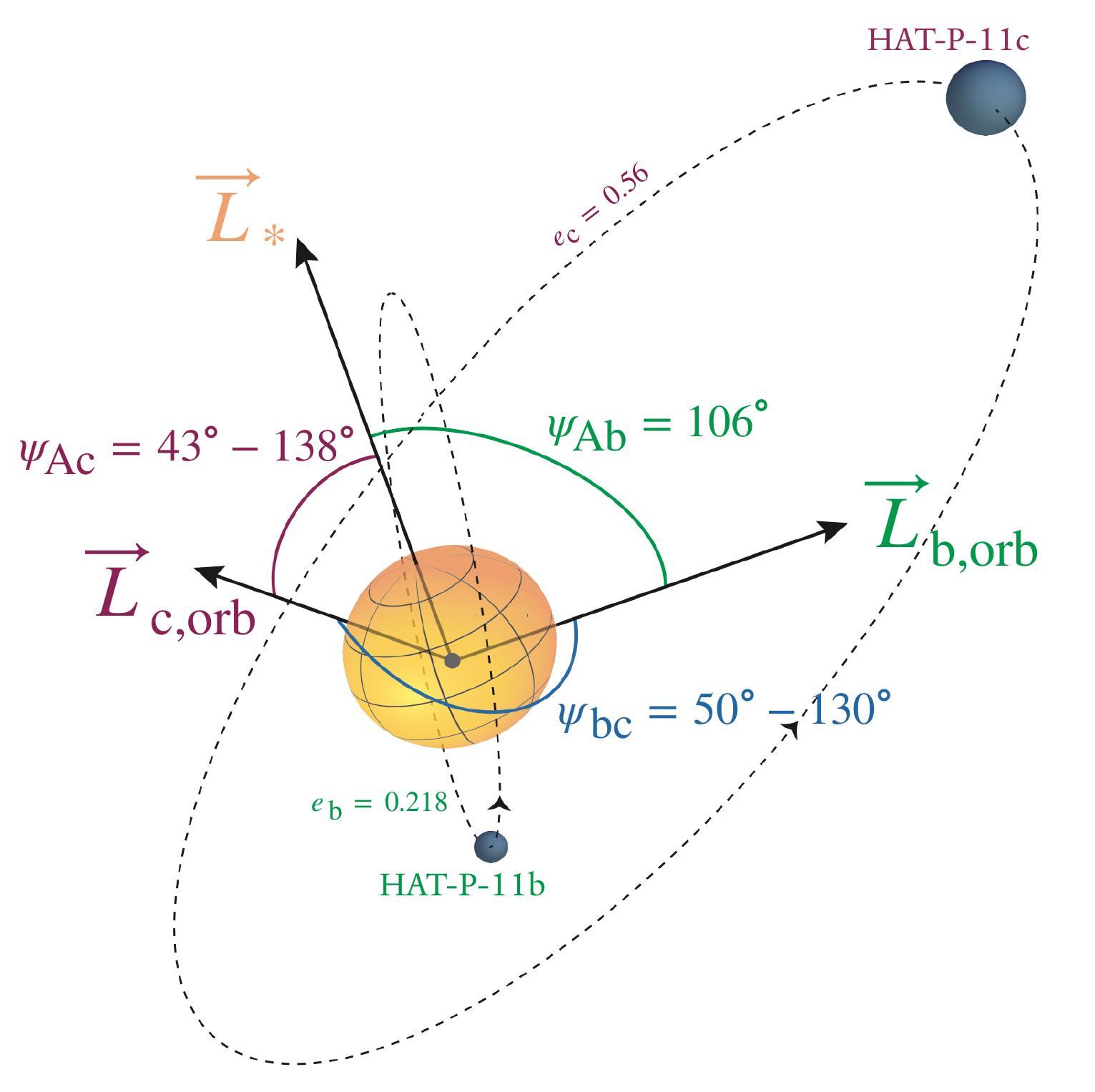}
    \caption{Schematic view of one possible orbital architecture of the HAT-P-11 system, based on posteriors from Table \ref{tab:params}. We label the eccentricities and misalignments between each orbit and the stellar spin axis.}
    \label{fig:hatp11_arch}
\end{figure}

The final piece of puzzle, the misalignment between the orbits of HAT-P-11 b and HAT-P-11 c, was initially explored by \cite{2020MNRAS_hatp11bc_obliquity}, in which they derived a bimodal inclination distribution for HAT-P-11 c, providing evidence of a large mutual inclination between planet b and c. \cite{an2024significant} (hereafter Paper 1) combined astrometry data from Hipparcos and Gaia with radial velocity data from Keck/HIRES to constrain the orbit of HAT-P-11 c, including a more precise bimodal constraint on inclination. All relevant orbital quantities of the system are well-constrained and are presented in Table \ref{tab:params}. In Paper 1, the two inclination modes are treated separately, together with the projected stellar spin constrained by \cite{2011ApJ_Sanchis-Ojeda_hatp11_star_spin}, to calculate the alignment of all three main angular momentum vectors in the HAT-P-11 system. Paper 1 reports the first measurement of the spin-orbit misalignment of HAT-P-11 c, and significantly improved measurements of the mutual inclination between HAT-P-11 b and HAT-P-11 c. A schematic view of the orbits is shown in Figure \ref{fig:hatp11_arch}. We are now in a position to postulate the complete dynamical history of HAT-P-11.

\begin{deluxetable}{lcr}
\tablewidth{0pt}
\tablecaption{Parameters of the HAT-P-11 System \label{tab:params}}
\tablehead{
Parameter & 
Value &
Reference}
\startdata
\multicolumn{3}{c}{Host Star} \\ \hline
      Mass (M$_\odot$)  & $0.81^{+0.02}_{-0.03}$ & \cite{Bakos+Torres+Pal+etal_2010}  \\
      Radius (R$_\odot$) & $0.75^{+0.02}_{-0.02}$ &\cite{Bakos+Torres+Pal+etal_2010}\\
      Age (Gyr) & $6.5^{+5.9}_{-4.1}$ & \cite{2018AJ_Yee_hatp11_RVS}\\
      \hline
      \multicolumn{3}{c}{HAT-P-11 b\tablenotemark{a}} \\ \hline
      Mass (M$_\oplus$)  & $23.4^{+1.5}_{-1.5}$ & \cite{2018AJ_Yee_hatp11_RVS}  \\
      Radius (R$_\oplus$)  & $4.36^{+0.06}_{-0.06}$ & \cite{huber_2017}  \\
      a (AU) & $0.05254^{+0.00064}_{-0.00066}$ & \cite{2018AJ_Yee_hatp11_RVS} \\
      e & $0.218^{+0.034}_{-0.031}$ & \cite{2018AJ_Yee_hatp11_RVS}
      \\
      $\psi_\mathrm{Ab} (^\circ)$ & $106^{+15}_{-14}$ & \cite{2011ApJ_Sanchis-Ojeda_hatp11_star_spin}\\
      \hline \multicolumn{3}{c}{HAT-P-11 c} \\ \hline
      Mass (M$_\mathrm{J}$)  & ${2.68\pm0.41}$ & Paper 1  \\
      a (AU) & $4.10\pm0.06$ & Paper 1 \\
      e & $0.652\pm0.017$ & Paper 1 \\
      $\psi_\mathrm{Ac} (^\circ)$ & 45 to 138\tablenotemark{b}
      & Paper 1\\ 
      $\psi_\mathrm{bc} (^\circ)$ & 49 to 131\tablenotemark{b}
      & Paper 1\\ 
\enddata
    \tablenotetext{a}{\cite{hirano_2011} and \cite{huber_2017} give slightly different values for spin-orbit misalignment and eccentricity, respectively.}
    \tablenotetext{b}{Range enclosing 95\% of the (highly non-Gaussian) posterior}
\end{deluxetable}

Both planets in the HAT-P-11 system represent dynamical puzzles, and the formation history of the system has been explored by \cite{petrovich_2020, Pu_Lai_2021} among others. In this work we propose a two-step formation mechanism: we posit that both HAT-P-11 b and HAT-P-11 c initially formed in circular aligned orbits, along with two other large planets. Through dynamical instability and subsequent planet-planet scattering, both additional planets were ejected from the system. As a result, HAT-P-11 b and HAT-P-11 c were left on eccentric and misaligned orbits, which triggered von-Zeipel-Lidov-Kozai (ZLK) migration in HAT-P-11 b. Through these mechanisms, all aspects of the present-day configuration of the system are reproduced.

The paper is organized as follows. In Section \ref{sec:scattering} we show that planet-planet scattering results in significant eccentricities and orbital misalignments for both planets, and we reproduce the present-day orbit of HAT-P-11 c. In Section \ref{sec:zlk} we similarly reproduce the present close-in state of HAT-P-11 b using a series of \textit{N}-body ZLK simulations accounting for both tidal and thermally driven radius evolution. We discuss implications of our study and conclude in Section \ref{sec:conclusions}.

\section{Planet-Planet Scattering}
\label{sec:scattering}
The high eccentricity of HAT-P-11 c, coupled with the significant spin-orbit misalignment reported in Paper 1, is the first dynamical puzzle. Naively, one would expect planets to form in circular, aligned orbits, thanks to strong eccentricity damping within the protoplanetary disk. Any deviation from this paradigm is believed to be the imprint of a dynamically active history. A number of well-understood mechanisms are known to excite the eccentricities and inclinations of planets, including but not limited to the von-Zeipel-Lidov-Kozai (ZLK) effect \citep{von_zeipel_1910, lidov_1962, kozai_1962} and secular chaos \citep{wu_2011}. Both of these mechanisms require an additional undetected companion in the system, something for which there is no direct evidence in the HAT-P-11 system.  The existing radial velocity data are satisfactorily explained by two planets with no residual trend, and the measured astrometric acceleration is similarly consistent with the two known planets (Paper 1).  To evade these constraints, an additional planet would have to be low-mass and/or very widely separated and hence be dynamically insignificant. Planet-disk interactions during planetary migration can excite eccentricities without the need for a companion, but this mechanism cannot generate the spin-orbit misalignment. 

In the absence of a more distant perturber, the most promising pathway is planet-planet scattering, where close encounters between pairs of planets lead to strong gravitational interactions that produce large eccentricities and obliquities. While additional planets would be required for this mechanism, violent scattering is prone to ejecting planets or scattering them to wide orbits where they would be difficult to detect \citep[e.g.,][]{chatterjee2008dynamical, carrera2019planet, Frelikh_2019}. In this section, we explore the generation of HAT-P-11 c's significant eccentricity and orbital misalignment through \textit{N}-body simulations of planet-planet scattering.

\subsection{Theoretical Background}
A wealth of literature exists on the subject of planet-planet scattering. Analytic treatment of this problem is difficult, so most analyses have been performed with numerical simulations. For a comprehensive review, see \cite{davies_2004}. Here we summarize some of the most relevant key results for our study. 

The stability of systems involving only two planets is well-understood analytically \citep{Gladman93}. Studies such as \cite{Ford_2001}, \cite{petrovich_2014} and \cite{gratia_fabrycky_2017} have shown that two-planet systems can attain a wide range of eccentricities but very rarely high mutual inclinations. Three-body scattering, on the other hand, can both excite eccentricity to arbitrarily large values and raise inclination up to $90^\circ$ \citep{chatterjee2008dynamical, juric_tremaine_2008, carrera2019planet}. In the spin-orbit misalignment posterior derived in Paper 1, the $95\%$ confidence interval is given by $\ang{45}$ to $\ang{138}$; thus the $\sim$40$^\circ$ lower limit requires three-body scattering to be invoked.

The stability of three-body systems cannot be obtained analytically. From numerical studies, system stability is both highly chaotic and dependent on mean-motion resonances \citep[e.g.,][]{Marzari_2014, Rath22}. However, to first-order the time to dynamical instability grows logarithmically with mutual separation \citep{marzari_2002}. The mutual separation is commonly parameterized by the mutual Hill radius (though note that other criteria often offer better predictions, e.g \cite{lammers_2024}):
\begin{equation}
    R_H = \frac{a_{i} + a_{i-1}}{2} \left(\frac{m_{i} + m_{i-1}}{3 M_*}\right)^{1/3} ,
\end{equation}
where $a_i$, $m_i$ are the semimajor axes and masses of the $i$th planet and $M_*$ is the mass of the star. The $\Delta$ parameter denotes the separation in units of the mutual Hill radius, 
\begin{equation}
    \Delta = \frac{a_{i} - a_{i-1}}{R_H}.
\end{equation}
The instability timescale grows as the separation between the planets increases \citep[e.g.,][]{Chambers_1996, Pu15, Tamayo20}.

\subsection{N-Body Simulations}
Scattering simulations are computationally expensive, and there is a large parameter space that the primordial system could inhabit. To work around this, we adopt a sequential approach.

We first use three-planet simulations, with HAT-P-11 c and two additional bodies but without HAT-P-11 b, to show that planet-planet scattering can exactly reproduce HAT-P-11 c's present-day orbit with the other bodies either ejected from the system or scattered to very wide orbits.  This step assumes that we may neglect HAT-P-11 b for the purposes of reproducing HAT-P-11 c's observed parameters.  We test this assumption by running an additional suite of scattering simulations with HAT-P-11 b present on an initial 0.3 AU orbit (roughly five times its present-day semimajor axis). In this suite of simulations with HAT-P-11 b, we do not seek to exactly reproduce HAT-P-11 c's orbit, given the extremely tight constraints on its semimajor axis and the large chaotic parameter space introduced by adding another surviving planet to the system. Rather, we content ourselves with showing that these simulations qualitatively match the architecture expected from the three-planet scattering simulations.

Our tiered approach to N-body simulations allows us to construct a statistical picture of the likely configurations of HAT-P-11 system after violent scattering from an initial four-planet configuration.  We can then estimate the fraction of the available parameter space that can produce the high mutual inclinations and significant eccentricities that are observed for HAT-P-11 c, along with a range of plausible initial conditions for HAT-P-11 b's subsequent orbital evolution.  This evolution under the influence of ZLK oscillations and tides is discussed in Section \ref{sec:zlk}.

\subsubsection{Three-Planet Simulations}
We perform 1000 three-planet simulations using the \texttt{IAS15} integrator \citep{rein_ias15} in the \reboundcodename \textit{N}-body integrator package \citep{Rein_2012}. We use the adaptive timestep criterion described in \cite{pham_2024}. The setup of our simulations is as follows: we consider three identical planets with $m_1 = m_2 = m_3 = 2.68 \: M_\mathrm{J}$, the best-fit value for HAT-P-11 c's mass. We note that our results are not sensitive to the masses of the three planets. We also ran a set of simulations with $m_1 = m_\mathrm{c}, m_2 = m_\mathrm{c}/2$ and 
$m_3 = m_\mathrm{c}/4$ following the prescription of \cite{anderson_2020}, and found no significant differences at the population level.

Three-body scattering tends to result in a single surviving planet, or additional survivors on highly eccentric/marginally unbound orbits \citep[e.g][]{carrera2019planet}. In both cases, the orbital energy of the non-innermost planets is nearly zero. Conservation of energy thus demands
\begin{equation}
    \frac{1}{a_f} = \frac{1}{a_{1,i}} + \frac{1}{a_{2,i}} + \frac{1}{a_{3,i}}.
\end{equation}
We set $a_f = 4.10$ AU, the present-day semi-major axis of HAT-P-11 c. In the interest of minimizing computation time, we set $\Delta = 3$ to induce rapid instability and scattering in our systems. This is unrealistically compact for primordial planetary systems, where separations would be expected to be around 5-10 mutual Hill radii \citep{raymond2022planet}. Changing the mutual separation of the planets does not qualitatively affect the scattering outcome, only the timescale until dynamical instability \citep[e.g][]{chatterjee2008dynamical, anderson_2020}. The above criterion hence informs the initial semi-major axes of the three planets: $a_{1,i} = 8.759$ AU, $a_{2,i} = 12.930$ AU, $a_{3,i} = 19.086$ AU. In each simulation, the other relevant orbital parameters for each planet are chosen at random from the following uniform distributions:
\begin{equation}
    \begin{split}
        e & \sim \mathcal{U}(0.01, 0.05)\\
        i & \sim \mathcal{U}(0^\circ, 2^\circ)\\
        f, \omega, \Omega & \sim \mathcal{U}(0, 2 \pi)\\
    \end{split}
\end{equation}

Note that we do not track the motion of the stellar spin axis in our simulations. Rather, we assume that the star's initial spin axis points along the $z$-axis and does not evolve significantly over the course of our simulations. With this assumption, the orbital inclination $i$ can be used as a proxy for the spin-orbit misalignment $\psi_\mathrm{Ac}$. We account for collisions via merging planets (conserving mass, volume and momentum but not energy) using the \texttt{reb\_collision\_resolve\_merge} collision module. We also remove particles if they exceed a distance of $10^3$ AU from the origin, using the \texttt{exit\_max\_distance} feature.

\begin{figure}
    \centering
    \includegraphics{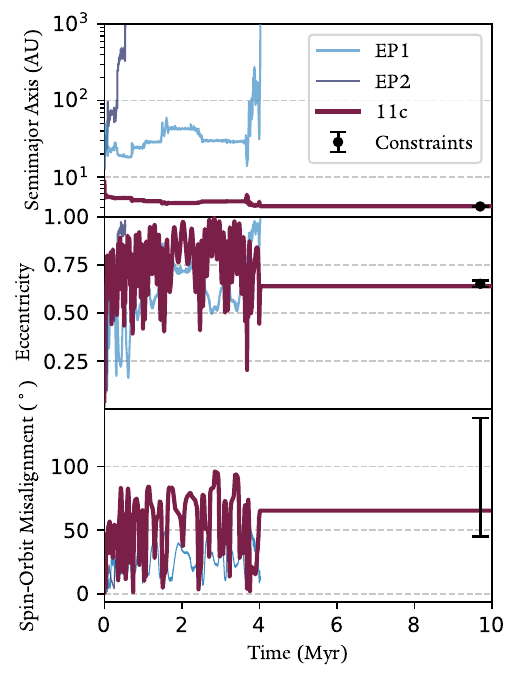}
    \caption{Example three-body scattering results for the origin of HAT-P-11 c. The thick maroon line and the two thin blue lines represent the orbital elements of HAT-P-11 c and the two ejected planets (EP), respectively. The subplots show the semimajor axis, eccentricity, and spin-orbit misalignment evolution of the three bodies, from top to bottom. When a planet is fully ejected from the system (heliocentric distance $>10^3$ AU) or becomes unbound $(a < 0)$, the line associated with the evolution of that planet is terminated. The black dot on the right-hand side of each subplot denotes our derived values for the present-day system, with the error bar representing the $1\sigma$ posteriors for semimajor axis and eccentricity and the $2\sigma$ posterior for spin-orbit misalignment.}
    \label{fig:3body}
\end{figure}

Figure \ref{fig:3body} shows an example of one of our simulations. The system rapidly enters a regime of dynamical instability, as expected from the initialization of the planets at small mutual Hill radii. This epoch is characterized by sudden perturbations in semimajor axis, eccentricity and inclination triggered by close encounters. Within 1 Myr, one planet is completed ejected from the system. The remaining two planets continue to experience perturbations in eccentricity and inclination, until a second planet is ejected from the system around $5$ Myr. At this point, HAT-P-11 c has gained significant eccentricity and inclination. The final orbital elements of HAT-P-11 c are $a_\mathrm{c} = 4.12$ AU, $e_{\rm c} = 0.639$, and $\psi_\mathrm{Ac} = 65.3^\circ$, which are all time-averaged values over the last $5$ Myr of the simulation.

\begin{figure*}
    \centering
    \includegraphics{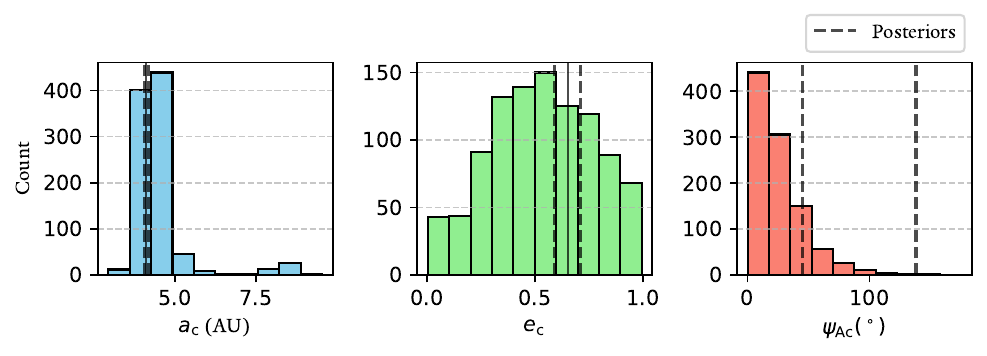}
    \caption{Results from our ensemble of three-body scattering simulations. We plot distributions in $a$, $e$ and $\psi_{\rm Ac}$ for the innermost planet in our three-planet simulations. The black lines correspond to the best-fit present-day values of HAT-P-11 c from Table \ref{tab:params}, with the dotted lines representing the $1\sigma$ error bars for $a_\mathrm{c}$ and $e_{\rm c}$ and $2\sigma$ for $\psi_{\rm Ac}$. All three orbital element ranges are well-represented in our sample, showing that this system can easily be produced through planet-planet scattering without fine-tuning.}
    \label{fig:scattering_stats}
\end{figure*}

In Figure \ref{fig:scattering_stats} we report statistics from the entire ensemble of simulations. We plot histograms of the distributions in $a_\mathrm{c}$, $e_{\rm c}$ and $\psi_{\rm Ac}$ of the innermost surviving planet in each simulation, as well as the present-day $1\sigma$ constraints ($2\sigma$ for $\psi_{\rm Ac}$) on the orbit of HAT-P-11 c from Paper 1. The simulation results are as expected: $a_\mathrm{c}$ is strongly peaked near the present-day value, and both $e_{\rm c}$ and $\psi_\mathrm{Ac}$ show a wide range of values that easily encompass the constraints. This shows that our system setup is capable of consistently producing orbits similar to that of HAT-P-11 c.

Of the $1000$ simulations we ran, the number with $1$, $2$ and $3$ surviving planets were $197, 802$ and $1$, respectively. The most common outcome by far is a system with two surviving planets, which would imply an external undiscovered planet in the HAT-P-11 system. 

The extra planet is not in tension with observations, because the extraneous surviving planet is generally very far out and hence would be undetectable. In some cases the extra planet produces long-term secular oscillations in the orbital elements of HAT-P-11 c, but this does not impact the system's stability. In our simulations there are a total of $908$ extraneous planets that survive. The $1\sigma$ semimajor axis distribution of these planets is $a_\mathrm{extra} = 72^{+123}_{-37}$ AU, while the perihelion distribution of these $a_{p, \mathrm{extra}} = 28^{+17}_{-9}$ AU. There are only 5 simulations where there is an orbit crossing between any of the extra planets and the surviving HAT-P-11 c. We do not analyze the stability of the surviving systems in-depth; an analytic criterion such as the one introduced by \cite{hadden_lithwick_2018} could be used to this end. We conclude that planet-planet scattering in a three planet system well-reproduces the unusual spin-orbit misalignment and eccentricity of HAT-P-11 c.

\subsubsection{Four-Planet Simulations}
\label{sec:fourbody}
We now consider the effect of planet-planet scattering on the orbit of HAT-P-11 b by running a suite of four-planet scattering simulations. The setup of these simulations is as follows. We perform 10,000 simulations using the new hybrid integrator \texttt{TRACE} \citep{Lu_TRACE} in \reboundcodename. We initialize the outer three bodies as in the previous section. However, we now initialize HAT-P-11 b on an $a_\mathrm{b} = 0.3$ AU orbit (roughly $5
\times$ its present-day semimajor axis), with all other orbital elements randomized in the same fashion as the three outer planets. The semimajor axis of HAT-P-11 b does not change significantly for most of our simulations. We adopt a timestep of $0.022$ years, which is roughly $1/9$th of HAT-P-11 b's initialized orbit. These simulations are also integrated for $10^7$ years. We also extend the maximum simulation distance to $10^4$ AU. The mean fractional energy error over our 10,000 simulations is $10^{-1.43}$.

\begin{figure}
    \centering
    \includegraphics{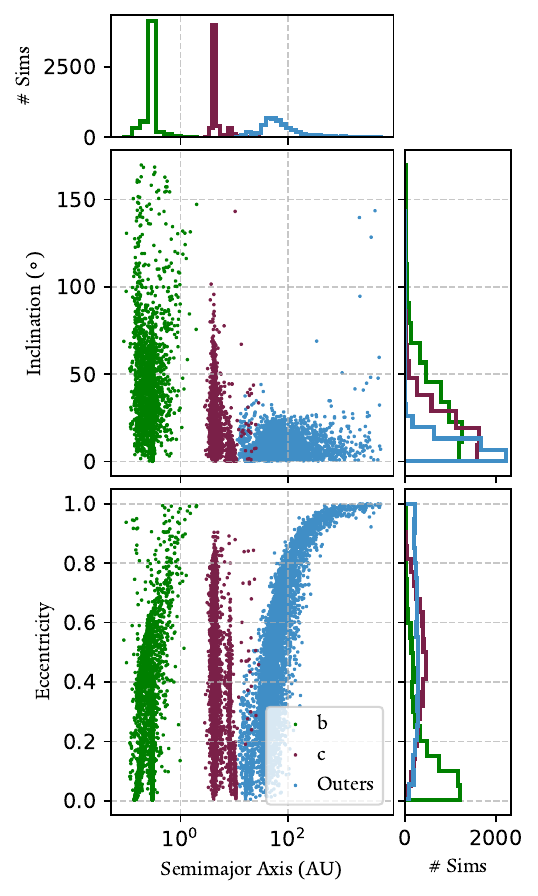}
    \caption{Results from an ensemble of 10,000 four-planet scattering simulations. We plot the eccentricities and inclinations of the planets representing HAT-P-11 b (green), HAT-P-11 c (maroon) and external planets (blue) as a function of semimajor axis. The points referring to HAT-P-11 b and HAT-P-11 c always correspond to the innermost planet and the innermost giant planet in our simulations, respectively. In every simulation these are distinct. The distributions of HAT-P-11 c and the outer planets are qualitatively consistent with our results from the three-planet scattering simulations. We see that HAT-P-11 b may be scattered to arbitrarily high eccentricities and very high inclinations.}
    \label{fig:4body}
\end{figure}

Figure \ref{fig:4body} shows statistics of orbital elements from our simulations. We restrict our discussion to simulations in which HAT-P-11 b survives on an orbit within $a_\mathrm{b} < 4$ AU; $5355$ simulations satisfy this criterion. Of these simulations, the vast majority ($4743$, or $89\%$) result in three-planet systems comprised of HAT-P-11 b, HAT-P-11 c, and a very distant outer perturber. This is consistent with our results from the previous subsection, which shows that HAT-P-11 b does not play a significant role in the scattering dynamics of the giant planets. $611$ simulations resulted in two-planet systems and $1$ simulation resulted in a four-planet system. We plot the eccentricities and inclinations of all planets corresponding to HAT-P-11 b, HAT-P-11 c and extraneous outer planets in their respective simulations as a function of semimajor axis. The semimajor axis $1\sigma$ distributions are given by $a_\mathrm{b} = 0.29_{-0.06}^{+0.00001}$ AU, $a_\mathrm{c} = 4.31_{-0.24}^{+0.40}$ AU and $a_\mathrm{outer} = 59_{-24}^{+78}$ AU. We also see that HAT-P-11 b can be scattered onto a wide range of inclinations and eccentricities. These $1\sigma$ distributions are given by $e_\mathrm{b} = 0.11^{+0.28}_{-0.07}$ and $\psi_\mathrm{Ab} = 25_{-17}^{+27}$ degrees. We also ran simulations initializing additional planets between the orbits of HAT-P-11 b and HAT-P-11 c, which are overwhelmingly ejected.

We conclude that the planet-planet scattering which was likely responsible for the eccentric and misaligned orbit of HAT-P-11 c also likely resulted in HAT-P-11 b gaining an eccentric and misaligned orbit. Crucially, this creates significant misalignment between the orbits of the two planets.

\section{ZLK Migration}
\label{sec:zlk}
The orbit of HAT-P-11 b is an even bigger puzzle than planet c. Its near-polar orbit with $\psi_{Ab} \sim 106^\circ$ \citep{2017AJ_hatp11_b_inclination} is highly unusual. Most intriguingly, planets as close-in as HAT-P-11 b would naively be expected to have perfectly circular orbits, due to significant tidal effects acting quickly to circularize the orbit. Indeed, the vast majority of hot Jupiters observed do have circular orbits \citep{dawson_johnson}. To first order, the tidal circularization timescale is given by \citep{goldreich_soter_1966}:

\begin{equation}
    t_\mathrm{circ} = \frac{4}{63}\frac{a^{13/2}}{\sqrt{G M_*^3}} Q_p m_p R_p^{-5}.
    \label{eq:tcirc}
\end{equation}
Plugging in the present-day values of the system and $Q_p = 10^5$, a fiducial estimate for the tidal quality factor of a Neptune-like planet \citep[e.g][]{millholland_2019}, we obtain $t_\mathrm{circ} = 2 \times 10^9$ years, which is compatible with the system's age of $6.5^{+5.9}_{-4.1}$ Gyr \citep{2018AJ_Yee_hatp11_RVS}. However, these expressions must be used with caution. \cite{wisdom_2008} found that the standard analytic expressions can underestimate tidal dissipation by several orders of magnitude for high eccentricities, and expressions such as Equation \eqref{eq:tcirc} significantly overpredict the circularization timescale. Given this, and the fact that tides act more efficiently at the highest eccentricities, we naively expect that HAT-P-11 b should be circularized. 

However, we cannot state with certainty that the orbit should be circularized. First, we note that the default calculation of $t_\mathrm{circ}$ is not that far off from the system's age. In addition, the unconstrained tidal quality factor $Q_p$ may well be higher than the nominal value of $10^5$ as argued by \cite{mardling_2008}. In this case, HAT-P-11 b could have formed at its present-day semimajor axis and gained significant eccentricity/spin-orbit misalignment through scattering alone, without having sufficient time to fully circularize. Given the timescales and assumptions enumerated, we consider this scenario a somewhat unlikely but certainly possible scenario.

In this section, we explore a more likely alternative, where ZLK oscillations (driven by HAT-P-11 c) and subsequent tidal circularization and migration act as an avenue to explain both the eccentricity and spin-orbit misalignment of HAT-P-11 b. We use initial conditions consistent with our four-body scattering simulations.

\subsection{ZLK Oscillations}
\cite{lidov_1962} and \cite{kozai_1962} independently identified a fascinating behavior characteristic of hierarchical three-body systems. Consider such a system, and denote the three bodies as the \textit{star}, \textit{b}, and \textit{c}. If the mutual inclination between the orbits of b and c are significantly misaligned  ($39.2^\circ < \psi_\mathrm{bc} < 140.8^\circ$), then the orbit of b will experience high-amplitude coupled oscillations in eccentricity and inclination in the absence of short-range forces such as general relativity. Commonly referred to as the Kozai-Lidov effect, the initial discovery of this mechanism by \cite{von_zeipel_1910} has been recently brought to light by \cite{ito_2019}. We hence adopt the name von Zeipel-Lidov-Kozai, or ZLK effect.

For an in-depth review of the ZLK effect as well as its many applications, see \cite{naoz_2016}. We provide in this paper a brief overview of the analytic understanding of the ZLK effect necessary to interpret our results. The simplest way to investigate the ZLK effect analytically is in the hierarchical secular approximation, which averages over the mean motion of both b and c \citep{ford_2000}. The validity of this approximation holds as long as $a_\mathrm{c} \gg a_\mathrm{b}$. The initial parameter space of the HAT-P-11 system in our simulations does not always satisfy this requirement ($a_{\rm b}/a_{\rm c} <0.1$), so we will use direct \textit{N}-body simulations to explore our system to obtain more accurate results. However, good first-order understanding can still be gained from the secular approximation\footnote{The extent to which the secular approximation is valid for mildly hierarchical systems is discussed in \cite{grishin_2018}.}. In this approximation and assuming b is a test particle, c's orbit remains unchanged throughout the system's evolution, and b's semimajor axis remains fixed. The maximum amplitude of the eccentricity oscillations depends on the initial mutual inclination: assuming the inner begins on a circular orbit, $e_{\mathrm{b, max}} = \sqrt{1 - (5/3) \cos^2 \psi_\mathrm{bc}}$ (in the quadrupole limit). The period of the ZLK oscillations is of order \citep[e.g.,][]{fabrycky_tremaine}:

\begin{equation}
\label{eq:zlk_timescale}
    \tau_{\rm ZLK} = \frac{2 P^2_{\rm c}}{3 \pi P_{\rm b}} \frac{m_\mathrm{star} + m_{\rm b} + m_{\rm c}}{m_{\rm c}} (1 - e^2_{\rm c})^{3/2},
\end{equation}
where $P_\mathrm{b}, P_\mathrm{c}$ are the periods of the inner and outer bodies, respectively.

The genesis of these effects is the apsidal precession of b's orbit, $\dot{\omega}_{\rm b}$, generated by the weak perturbations of the outer. In the secular approximation, the apsidal precession rate can be explicitly calculated. The orbit-averaged Hamiltonian of the system is \citep{ford_2000, fabrycky_tremaine}:

\begin{equation}
    \begin{split}
    \langle \mathcal{H} \rangle = & -\frac{G m_\mathrm{star} m_{\rm b}}{2 a_\mathrm{b}} - \frac{G(m_\mathrm{star} + m_{\rm b}) m_{\rm c}}{2 a_\mathrm{c}} \\
    & - \frac{G m_\mathrm{star} m_{\rm b} m_{\rm c}}{m_\mathrm{star} + m_{\rm b}} \frac{a_\mathrm{b}^2}{8 a_\mathrm{c}^3 (1 - e_{\rm c}^2)^{3/2}} \\
    & \times \big[ 2 + 3 e_\mathrm{b}^2 - \sin^2 \psi_\mathrm{bc} (3 + 12 e_\mathrm{b}^2 - 15 e_\mathrm{b}^2 \cos^2 \omega_\mathrm{b}) \big]
    \end{split}
\end{equation}
where $\psi_\mathrm{bc}$ is the mutual inclination between bodies $b$ and $c$. This equation can be solved for the apsidal precession rate; see \cite{naoz2013, naoz_2016} for an in-depth derivation:

\begin{equation}
\label{eq:omegazlk}
\begin{split}
    \dot{\omega}_{
\rm b, \mathrm{ZLK}} = 6 C_2 \bigg[& \frac{1}{G_{\rm b}} \big(4 \cos^2 \psi_\mathrm{bc} + (5 \cos(2 \omega_\mathrm{b}) - 1) \\
    & \times ( 1 - e_\mathrm{b}^2 - \cos^2 \psi_\mathrm{bc}) \big) \\
    & + \frac{\cos \psi_\mathrm{bc}}{G_{\rm c}} \big(2 + e_\mathrm{b}^2 (3 - 5 \cos(2 \omega_\mathrm{b})\big)\bigg]
\end{split}
\end{equation}
where 
\begin{equation}
    C_2 \equiv \frac{G^4}{16} \frac{(m_{\rm s} + m_{\rm b})^7}{(m_{\rm s} + m_{\rm b} + m_{\rm c})^3} \frac{m_{\rm c}^7}{(m_{\rm s}m_{\rm b})^3} \frac{L_{\rm b}^4}{L_{\rm c}^3 G_{\rm c}^3}
\end{equation}
where $G$ is the gravitational constant, $L_\mathrm{b/c}$ are the Delaunay action variables, and $G_\mathrm{c} \equiv L_\mathrm{c} \sqrt{1 - e_\mathrm{c}^2}$.
Note that this level of approximation is only strictly valid in the case of a circular outer orbit and an inner test particle, neither of which are good assumptions in our system. The next level of approximation, the octupole-level, would correctly account for both of these complications. However, the dynamics are chaotic in the octupole order \citep{Li14}. As we aim only to draw first-order analytical understanding from the secular model (to eventually be confirmed with full \textit{N}-body simulations), we work for now in the quadrupole approximation for greater clarity into the most impactful quantities in this problem.

ZLK oscillations will persist so long as the dominant source of precession in $\omega_\mathrm{b}$ is due to perturbations from c. As $a_\mathrm{b}$ becomes smaller due to tidal migration, a host of other effects become significant and may suppress the precession from c, and thus the ZLK oscillations themselves. For instance, the precession due to general relativity can be quantified \citep{eggleton_2001, fabrycky_tremaine}:

\begin{equation}
    \dot{\omega}_{\rm b, \mathrm{GR}} = \frac{3 G^{3/2} (m_\mathrm{star} + m_{\rm b})^{3/2}}{a_\mathrm{b}^{5/2} c^2 (1- e_\mathrm{b}^2)}
\end{equation}
where $c$ here is the speed of light, not to be confused with a quantity associated with body c. \cite{2018AJ_Yee_hatp11_RVS} ruled out the possibility of ZLK oscillations being active in the present-day HAT-P-11 system, on the basis of general relativity suppressing ZLK precession at HAT-P-11 b's present-day semimajor axis. Plugging in values associated with the present-day system, we confirm that ZLK oscillations are expected to be quenched by GR precession, with $\dot{\omega}_{\rm b, \mathrm{ZLK}} / \dot{\omega}_{\rm b, \mathrm{GR}} \sim 0.8$. Hence, HAT-P-11 b's large eccentricity cannot be attributed to it \textit{currently} being in a high-eccentricity phase of a ZLK oscillation. However, this does not rule out the possibility of ZLK oscillations having occurred in the past. In order for this to have been the case, HAT-P-11 b must have originated at a larger semimajor axis.

\subsection{Tidal Friction}
\label{sec:tid}
We turn to the process of ZLK migration, which has been invoked to explain the anomalously high eccentricity and tight orbits of a number of hot Jupiters (see \cite{wm2003}, \cite{mardling_2010}, \cite{beust_2012} among others). The idea behind ZLK migration is to combine the concepts of ZLK oscillations and tidal evolution. During each high-eccentricity phase of the ZLK oscillation, tidal friction becomes significant due to b's close approach with the host star.

In this work, we will consider the prescription of equilibrium tides only \citep{hut_binary,eggleton1998equilibrium, mardling2002calculating}. In this framework, the strength of tidal friction is parameterized by the constant time lag $\tau$, which represents the time lag between the tidal bulge of a body and the line of centers connecting to the tidal perturber. $\tau$ may be related (with caution, see \cite{leconte2010tidal, Lu_2023}) to the commonly used tidal quality factor $Q$ via

\begin{equation}
    \tau = (2 n Q)^{-1}
\end{equation}
with $n$ the mean motion of the tidal perturber, in this case the host star. We note that more complex and nuanced tidal models are available. In particular, for the high-eccentricity ZLK epochs dynamical tides may be a more accurate prescription \citep[e.g][]{mardling_1995, lai_2012, fuller_2014, vick_2019, vick_misaligned_zlk}. However, no self-consistent \textit{N}-body package exists with these more complex prescriptions. Hence, we restrict ourselves to the simpler equilibrium tide model in the present work, though we encourage future works to explore the effect of more complex tidal models.

\begin{figure}
    \centering
    \includegraphics[width=0.48\textwidth]{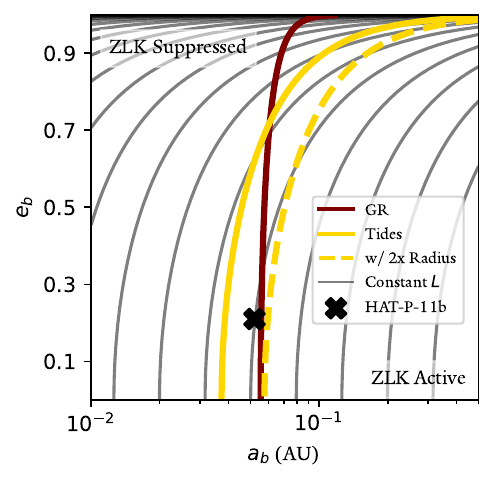}
    \caption{Comparison of precessions from the ZLK effect to general relativity and tides in semimajor axis-eccentricity space, using HAT-P-11's present-day configuration. The maroon line represents where the precessions due to general relativity and ZLK are exactly equal, while the gold lines represent tides with HAT-P-11 b's present day radius (solid) and tides with twice this radius (dashed). Lines of constant angular momentum are shown in gray. HAT-P-11 b's present-day position is shown as a black cross. A planet experiencing ZLK-induced high-eccentricity migration will move towards the upper left hand corner in eccentricity-semimajor axis space until it encounters one of the colored curves, at which point ZLK oscillations are quenched and it follows one of the contours of constant angular momentum to zero eccentricity.}
    \label{fig:ratio}
\end{figure}

In the equilibrium tide model, bodies are no longer point particles but rather are endowed with structure. The two improvements on the point particle model are the effect of a misaligned tidal bulge and the rotational flattening of each body. Both are significant to the dynamics of ZLK migration, not only because they act to tighten the orbit of b over time, but also because they both introduce additional precession terms. For our system, tidal precession is expected to be more relevant, so we will focus on that aspect. The precession rate associated with tides is given by \citep{fabrycky_tremaine}

\begin{equation}
\label{eq:omegatide}
\begin{split}
    \dot{\omega}_\mathrm{Tide} = &  \frac{15 \sqrt{G (m_\mathrm{star} + m_{\rm b})}}{8 a_\mathrm{b}^{13/2}} \frac{8+12 e_\mathrm{b}^2 + e_\mathrm{b}^4}{(1 - e_\mathrm{b}^2)^{9/2}} \\
    & \times \frac{1}{2}\bigg[\frac{m_{\rm b}}{m_\mathrm{star}} k_{\mathrm{star}} R_\mathrm{star}^5 + \frac{m_\mathrm{star}}{m_{\rm b}} k_{\rm b} R_{\rm b}^5\bigg],
\end{split}
\end{equation}
where $k_\mathrm{star}, k_\mathrm{b}$ are the tidal Love numbers of the star and planet, respectively. Figure \ref{fig:ratio} demonstrates the significance of including the effects of tidal friction. In this figure we consider a system with a central object consistent with HAT-P-11A and an outer perturber consistent with HAT-P-11 c. The semimajor axis and eccentricity of HAT-P-11 b are varied, and we analyze where in semimajor axis-eccentricity space ZLK oscillations are allowed. The present-day orbit of HAT-P-11 b is also marked. The maroon and gold lines mark where $|\dot{\omega}_\mathrm{ZLK}/\dot{\omega}_\mathrm{GR}|=1$ and $|\dot{\omega}_\mathrm{ZLK}/\dot{\omega}_\mathrm{Tide}|=1$, respectively. Towards the upper left (lower semimajor axis and higher eccentricity) of each line, ZLK oscillations are quenched by the respective additional effect, while towards the bottom right ZLK oscillations are allowed to be active. From Equation \eqref{eq:omegatide}, we see that $\dot{\omega}_\mathrm{Tide}$ scales strongly with the eccentricity of b. As a result, while tides are less influential than GR at low eccentricities, the opposite is true at higher eccentricities. This is highly impactful for ZLK oscillations: a system where b is at a sufficiently high semimajor axis will be able to sustain ZLK oscillations of extremely high amplitude if only GR is considered, but if tides are accounted for as well during a high-eccentricity epoch the ZLK effect will be quenched.

Consider now the evolution of a body under the influence of tidal effects experiencing ZLK oscillations. If octupole level effects are accounted for \citep{naoz2013, naoz_2016}, the maximum eccentricity of the ZLK oscillations is not constant; rather, the maxima increase slowly over the octupole timescale \citep{naoz2013, li_eccentricity_2014}. In addition, during each high-eccentricity epoch tides will shrink the orbit. The result is that over the course of ZLK oscillations, b's orbit will move towards the upper left corner of Figure \ref{fig:ratio}. Inevitably, at some point there will be a crossing (mostly likely with the gold line), and ZLK oscillations will be quenched. Then b is marooned on a high-eccentricity orbit. This state is only temporary, however, as tides continue to be significant. Now, tidal dissipation acts efficiently to circularize and shrink the planet's orbit, until b settles into a close-in, circular orbit. Tidal circularization conserves angular momentum $L_{\rm b}$, so b's evolution through phase space will follow the contour of constant $L_{\rm b}$ (plotted in grey) intersecting the gold line until the planet reaches zero eccentricity.

Taking this together, the basic idea of our hypothesis is as follows. HAT-P-11 b initially formed at a significantly greater semimajor axis than we see today. The violent scattering events described in the previous section resulted in a significant orbital misalignment between HAT-P-11 b and HAT-P-11 c, which initiated ZLK oscillations in HAT-P-11 b's orbit. Tidal dissipation gradually tightened HAT-P-11 b's orbit until tidal precession broke the ZLK oscillations, at which point the highly eccentric HAT-P-11 b began the process of tidal circularization. We propose that HAT-P-11 b is in this final state, having not yet fully circularized. This is similar to the story proposed by \cite{beust_2012} to explain the orbit of the similarly eccentric close-in sub-Neptune GJ 436, but in their case the outer perturber is not well constrained.

\subsection{Radius Evolution}
It turns out that the story we just elucidated is incomplete, since simple ZLK migration alone cannot explain HAT-P-11 b's orbit. This is due to how relatively close-in HAT-P-11 c is. From Equation \eqref{eq:omegazlk}, the precession rate associated with the ZLK effect scales very strongly with $a_\mathrm{c}$. Since $a_\mathrm{c}$ is small, the precession associated with ZLK oscillations is strong, and thus very high eccentricities are required for tides to break the ZLK cycles (e.g., in comparison to GJ 436). Inspection of Figure \ref{fig:ratio} provides a clearer picture. The contour that HAT-P-11 b presently lies on does not intersect the solid golden line in the phase space shown.  While it does intersect the dark red line denoting suppression of ZLK oscillations due to GR precession, this mechanism would require suppression of ZLK oscillations at an eccentricity of just $\approx$0.5 at a semimajor axis where tidal damping is slow, and is not something that we see in our simulations.  In order for HAT-P-11 b to have a constant $L$ path from tidal quenching of ZLK oscillations to its current orbit, we would have to extend Figure \ref{fig:ratio} well to the right. We find that initializing HAT-P-11 b at these larger orbital separations typically results in dynamical instability, and often ejection of HAT-P-11 b. 
In other words, the quenching of ZLK by tidal effects in a reasonable initial parameter space cannot bring HAT-P-11 b to its present-day orbit via tidal circularization. We see that more realistic contours of constant $L$ terminate significantly closer to the star, sometimes by up to an order of magnitude. This tendency to overshoot HAT-P-11 b's present-day orbit is insensitive to initial conditions and is confirmed with \textit{N}-body simulations. Thus, the present-day configuration cannot be explained with traditional ZLK migration with fixed planet radius and equilibrium tides alone. 

However, the planet's radius can be inflated due to tidal heating during ZLK migration. Radius evolution of planets due to tidal heating is a degree of freedom that is typically unexplored in studies of the ZLK effect. During ZLK oscillations, HAT-P-11 b will reach very high eccentricities and hence experience many close pericenter approaches with the host star. During these close approaches, tidal heating becomes very significant and acts to inflate the planet's envelope. Radius inflation from tidal heating is a well-explored concept: it has been invoked to explain the anomalously large radii of hot Jupiters \citep{blm_01} and sub-Saturns \citep{millholland_2019, sm_20}. These works show that the radius of a sub-Neptune or sub-Saturn can reasonably be inflated over a factor of two via this mechanism.

This effect may play a significant role in the dynamics of the system, as from Equation \eqref{eq:omegatide} tidal precession scales as $\propto R_{\rm b}^5$. Indeed, we see from the dotted golden line in Figure \ref{fig:ratio} that increasing the radius of HAT-P-11 b by a factor of two breaks the ZLK oscillations significantly earlier, and now there does exist a reasonable contour of constant $L_{\rm b}$ that can deliver HAT-P-11 b to its present-day orbit. Coupled dynamical and thermal evolution of hot Jupiters were explored in the context of planet-planet scattering by \cite{rozner_2022,glanz_22}, who found that accounting for radius inflation significantly accelerated the formation and destruction of hot Jupiters and enhanced the production of warm Jupiters. \cite{petrovich_2015} investigated ZLK migration for initially inflated planets which slowly shrink down. However, fully self-consistent coupled dynamical and thermal evolution in the context of ZLK migration has not been explored in detail. Given the uncertainties in our system, we employ a simplistic model for the radius inflation of HAT-P-11 b in this work, and we defer a more nuanced prescription for self-consistent thermal and orbital evolution to future work.

\cite{sm_20} generated 10,000 \texttt{MESA} models \citep{paxton_2011}, which modify and build upon the publicly available \texttt{MESA} models for sub-Saturns developed by \cite{chen_rogers_2016}. These models spanned four principal parameters: mass $M$, envelope mass fraction $f_{\mathrm{env}}$, flux from the host star $F$, and internal luminosity $\mathcal{L}$\footnote{Note that \cite{sm_20} parameterized the tidal luminosity with the factor $\Gamma \equiv \log_{10} \left[\frac{Q' (1 + \cos^2 \epsilon)}{\sin^2 \epsilon}\right]$, where $Q' \equiv 3Q/2k_2$ is the reduced tidal quality factor and $\epsilon$ is the planetary obliquity. This is because both \cite{sm_20} and their progenitor study \cite{millholland_2019} focused on obliquity tides. However, we are agnostic to the specific source of tides in this study, and as such require only the total luminosity deposited at the core-envelope interface.}. Each model was then evolved forward in time for $10$ Gyr, after which a final radius $R$ was reported. We fit a subset of the \cite{sm_20} models appropriate to HAT-P-11 b in order to construct a luminosity-radius relationship using a fourth-order polylogarithmic function:

\begin{equation}
\label{eq:lr_interpolation}
    \begin{split}
    \frac{R}{R_\mathrm{E}} = & A \left[\log_{10} \left(\frac{\mathcal{L}}{\mathcal{L}_\odot}\right)\right]^4 + B \left[\log_{10} \left(\frac{\mathcal{L}}{\mathcal{L}_\odot}\right)\right]^3 \\
    & + C \left[\log_{10} \left(\frac{\mathcal{L}}{\mathcal{L}_\odot} \right)\right]^2 + D \left[ \log_{10} \left(\frac{\mathcal{L}}{\mathcal{L}_\odot}\right)\right] \\
    & + E
    \end{split}
\end{equation}
where $A = 5.23 \times 10^{-4}$, $B = 3.37 \times 10^{-2}$, $C = 0.801$, $D = 8.357$, and $E = 3.596 \times 10^1$. Here the orbit-averaged tidal luminosity of the planet is given by \citep{mardling2002calculating}:

\begin{equation}
\label{eq:l_averaged}
    \begin{split}
    \langle \mathcal{L} \rangle = & -\mu_{\rm b*} a_\mathrm{b}^2 n_{\rm b} \left(\frac{m_*}{m_{\rm b}}\right) \left(\frac{R_{\rm b}}{a}\right)^5 \left(\frac{3 k_{2,b}}{2 Q_{\rm b}}\right) \\
    & \times \bigg[\frac{1}{2} \left(\Omega_e^2 h_1(e) + \Omega_q^2 h_2(e)\right) + \Omega_h^2 h_3(e) \\
    & - 2 n_{\rm b} \Omega_h h_4(e) + n^2 h_5(e) \bigg],
    \end{split}
\end{equation}
where $\Omega_e, \Omega_h, \Omega_q$ are functions of the planetary spin axis and present orbital configuration, while $h_1, h_2, h_3, h_4, h_5$ are functions of eccentricity. The exact expressions are given in the Appendix. Note the dependence on both orbital parameters and the spin axis. Since our code self-consistently tracks the spin and orbital evolution, this expression correctly accounts for both eccentricity and obliquity tides.

\begin{figure}
    \centering
    \includegraphics[width=0.48\textwidth]{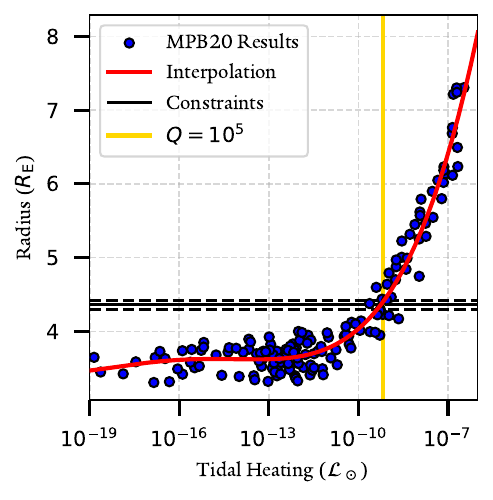}
    \caption{Radius-luminosity relationship derived from the models of \cite{sm_20}. In blue are 158 of their \texttt{MESA} model results using the cuts described in the main text. The red line represents our best-fit to the data, using a fourth-order polylogarithmmic function given in Equation \eqref{eq:lr_interpolation}. The gold vertical line shows the tidal heating at HAT-P-11 b's present-day orbital configuration given $Q = 10^5$. The black horizontal lines show the observational constraints on HAT-P-11 b's radius. Note that the gold and red lines intersect with the black line, meaning HAT-P-11 b's present-day radius is reproduced.}
    \label{fig:lr_relation}
\end{figure}

We select from the models generated by \cite{sm_20} as follows. Out of the 10,000 models they generated, we take a slice in mass ($20 \: \mathrm{M}_\mathrm{E} < \mathrm{M} < 25 \: \mathrm{M}_\mathrm{E}$) appropriate for the observational constraints on HAT-P-11 b. The value of $f_\mathrm{env}$ is not at all constrained, but the present-day radius and orbital parameters of HAT-P-11 b are well-constrained \citep{2018AJ_Yee_hatp11_RVS}. It is therefore vital to match these parameters at the end of our simulations. There is a degeneracy between $f_\mathrm{env}$ and the value of the tidal quality factor $Q$ in reproducing HAT-P-11 b's radius at its present-day orbit, since the observed radius could be equally well-described by a small envelope mass with efficient dissipation (low $Q$) or by a large envelope mass with weaker dissipation (high $Q$). We assume a fiducial value of $Q = 10^5$, and then use Equations \eqref{eq:lr_interpolation} and \eqref{eq:l_averaged} to match the present-day radius of HAT-P-11 b given its present-day orbital parameters. There is some degree of uncertainty in the precise values of many of the values in Equation \eqref{eq:l_averaged} at the end of our simulations, but we can make a number of simplifying assumptions to obtain a ballpark estimate. Most of the uncertainties lie in the values of $\Omega_e$, $\Omega_h$ and $\Omega_q$, which are dot products of the spin vector with the Runge-Lenz vector, orbit normal, and the cross product of the two former, respectively. If we assume zero planetary obliquity as expected given significant tidal dissipation \citep{su_lai_tidal}, the spin axis will be aligned with the orbital angular momentum, and $\Omega_e, \Omega_q = 0$. If we further assume that the planet rotates at the pseudo-synchronous spin rate, $\Omega_h$ can be calculated. Plugging these values and orbital parameters associated with HAT-P-11 b's present-day orbit, along with $Q = 10^5$, gives the present-day tidal luminosity of HAT-P-11 b. We select a range of $f_\mathrm{env}$ such that there is a good match with this value. We ultimately choose $0.05 \: < f_\mathrm{env} < 0.1$. With these slices we consider 158 models. The models we selected, as well as our best-fit line, are plotted in luminosity-radius space in Figure \ref{fig:lr_relation}.

We now use our fit to the \texttt{MESA} models to evolve HAT-P-11 b's radius during the simulations. In the interest of computation time and as our expression is orbit-averaged, we do not recompute the planet's radius every timestep. Rather, radius evolution is computed once per orbit of HAT-P-11 b. Once per orbit, the orbit-averaged luminosity is calculated and fed into our luminosity-radius relation Eq. \eqref{eq:lr_interpolation}, and the radius of HAT-P-11 b is correspondingly updated. We again emphasize that our prescription for the thermal radius evolution of the planet is a first-order approximation, where we assume the planet's radius responds effectively instantaneously to the tidal luminosity. A more realistic prescription would include a time-delayed response to the tidal heating, and a fully self-consistent numerical prescription could involve codes that couple the dynamical evolution of the system to the planet's thermal evolution. For example, see \cite{glanz_22} who coupled outputs from \texttt{MESA} \citep{paxton_2011} and \texttt{AMUSE} \citep{PZwart_2009}. While we do not use these nuanced prescriptions in the present work, we assert that our first-order approximation qualitatively achieves similar results. As our orbital code tracks tidal luminosity self-consistently, our planet will ultimately be inflated to the same degree, and the ZLK cycles will be broken at similar times. The improvement which would likely produce the largest discrepancy between our model is our choice of equilibrium tides, but we defer treatment of the more realistic dynamical tide model to future work.

\subsection{N-body simulations}
\label{sec:zlk_setup}
We investigate the present-day orbit of HAT-P-11 b via ZLK migration through a suite of \textit{N}-body simulations using \texttt{REBOUND}. The setup of our numerical simulations are as follows. We initialize HAT-P-11 A and c in their present-day orbital configurations. We select the best-fit values for the mass, semimajor axis and eccentricity of HAT-P-11 c from Table \ref{tab:params}. The spin-orbit misalignment of HAT-P-11 c is less constrained; we draw from a uniform distribution $\psi_\mathrm{c, i} \in \{\ang{33.3}, \ang{50.26}\}$, where the lower bound is selected from the $3\sigma$ confidence interval from the observational posteriors, and the upper bound from the $2\sigma$ confidence interval of our four-body scattering simulations. The asymmetric lower bound is selected to encompass more of the highly non-Gaussian observational posterior. The nodes are randomly distributed. HAT-P-11 b is initialized with orbital elements randomly drawn from uniform distributions informed from the $2\sigma$ distributions obtained in our four-body scattering simulations: $a_\mathrm{b} = \{0.154 \ \mathrm{AU}, 0.518 \ \mathrm{AU}\}$, $e_\mathrm{b} = \{0.01, 0.70\}$ and $\psi_\mathrm{b} = \{1.01^{\circ}, 97.0^{\circ}\}$. We initialize the mass with the best-fit value from Table \ref{tab:params}, and the radius with Equation \eqref{eq:lr_interpolation}.

We note that our scattering simulations often include an extra surviving distant planet. The impact of distant perturbers on the classic ZLK hierarchical configuration was studied by \cite{best_petrovich_2022}, who found that it generates chaotic spin-orbit evolution and makes achieving retrograde orbits easier, which would be favorable to our scenario. However, the first-order dynamics do not change, so we do not model it in our simulations. We also comment on the possibility of direct ZLK oscillations occuring between HAT-P-11 c and this surviving distant perturber. This is a fairly rare occurrence; only 653 simulations, or 14\%, result in a mutual inclination between HAT-P-11 c and the surviving outer planet greater than $\ang{40}$, so ZLK oscillations are rarely triggered. However, for these simulations where the prerequisite mutual inclination for ZLK cycles is reached, $|\dot{\omega}_{\mathrm{ZLK}}| > |\dot{\omega}_{\mathrm{GR}}|$ so ZLK cycles are in fact active. Nevertheless, the parameter space for a surviving external planet is large and unconstrained from observation, and it doesn't change the dynamics of the inner planets qualitatively. Thus, we do not model this outer planet in our ZLK simulations.

We also model the effects of general relativity and tides. We use \texttt{REBOUNDx} \citep{tamayo2020reboundx} to model both. For GR precession, we use the \texttt{gr} prescription \citep{anderson_gr}, appropriate for systems with a dominant central mass. We consider self-consistent spin, tidal and orbital evolution using the \texttt{tides\_spin} effect \citep{Lu_2023}, which provides the ability to evolve the evolution of the stellar spin axis for consistent spin-orbit misalignment tracking as well. For planet b, we use fiducial parameters of $k_2 = 0.5$, $I = 0.25$ appropriate for a Uranus-like planet \citep{yoder1995astrometric, lainey2016}, and we initialize its spin state as both pseudo-synchronized and aligned with the orbit. Planet c is considered a point particle, and the effects of structure are not accounted for. While evolution of the stellar spin axis is considered as part of the equilibrium tides model, spin-down due to stellar evolution,  which would occur during the early phases of our simulations \citep{bouvier_1997}, is not modeled. These effects are not expected to be significant for our simulations. Authors such as \cite{bolmont_2012} and \cite{faridani_2023, Faridani24} explored the effects of stellar spin evolution on planetary orbits and only found meaningful impacts for planets within $a = 0.05$ AU, well within the hypothesized original orbit of HAT-P-11 b.

Tides on the host star and planet c are not considered, but we set the value of planet b's tidal parameter to $\tau_{\rm b} = 10^{-5}$ years. This is an unrealistically high value of $\tau$, chosen to minimize computation time: direct \textit{N}-body simulations with a realistic $\tau$ value proved to be computationally infeasible. This is a common practice in numerical simulations \cite[e.g][]{bolmont_2015, Becker20}. Scaling $\tau$ does not change the qualitative behavior of the system. Neither the ZLK timescale nor the tidal precession rate, Equations \eqref{eq:zlk_timescale} and \eqref{eq:omegatide} respectively, depend on the value of the tidal parameter. The only dependence on $\tau$ is the rate of orbital energy drained from tides. Thus, our scaled-up value of $\tau$ serves only to drain more energy from the planet's orbit every ZLK cycle (therefore decreasing the semimajor axis more) and increase the rate at which HAT-P-11 b circularizes once the ZLK oscillations are ultimately quenched. These both only serve to linearly modify the timescale in which the planet reaches the final stage of evolution.


We now briefly discuss what realistic values of $\tau$ may be. While the values of tidal parameters in exoplanets are very poorly constrained, estimates of Neptune's $\tau \sim 10^{-8}$ years are based on the orbit of its satellites Proteus and Larissa \citep{zhang_neptune}. Therefore, the tidal timescale in the N-body simulations is a factor $\sim$10$^3$ longer than the simulation timescale. 

\subsection{Simulation Results}

\begin{figure*}
    \centering
    \includegraphics[width=0.9\textwidth]{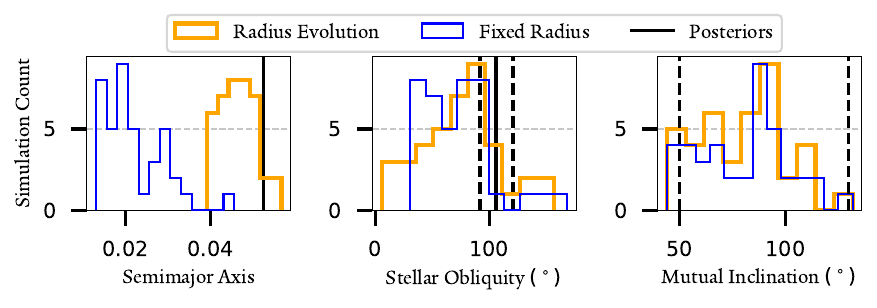}
    \caption{Results from a population synthesis of ZLK simulations. We consider $40$ simulations with radius evolution (orange) and $40$ with fixed radius (blue). We compare distributions of semimajor axis, spin-orbit misalignment and mutual inclination. We see that the fixed radius model cannot generate HAT-P-11 b's present-day semi-major axis, but the model with tidally-driven radius inflation can.}
    \label{fig:zlk_pop_synth}
\end{figure*}
Figure \ref{fig:zlk_pop_synth} shows distributions of spin-orbit misalignment and eccentricity for a population synthesis of $100$ systems. We initialize each simulation as described in the previous section and run until the ZLK cycles are quenched and the orbit has circularized to the present-day eccentricity $e_\mathrm{b} = 0.218$. At this point, the simulation is halted. To illustrate the effects of radius inflation, we also run the same initial condition but fix the radius of HAT-P-11 b to its present-day value throughout the course of the simulation. We allow each simulation to run for 10 Myr, which corresponds to 10 Gyr in when time is rescaled appropriately. $40$ simulations with radius inflation are able to quench ZLK oscillations and circularize in this timeframe, with the remaining $60$ either never initiating ZLK oscillations or being unable to quench them in time. For the simulations with fixed radius, $40$ simulations quench and circularize in time. For the population synthesis that considers radius inflation, we see that while the best-fit observational values are not the most favored there is nonetheless a high proportion of simulations that match the constraints, and we do not require fine-tuning to reproduce our posteriors. The population synthesis without radius inflation does not exhibit significant differences in the spin-orbit misalignment or mutual inclination distributions. However, as expected this model greatly underpredicts the semimajor axis of HAT-P-11 b: all simulations predict an orbit with $a_\mathrm{b} < 0.045$ AU (consistent with the implication of Figure \ref{fig:ratio}).

\begin{figure*}
    \centering
    \includegraphics{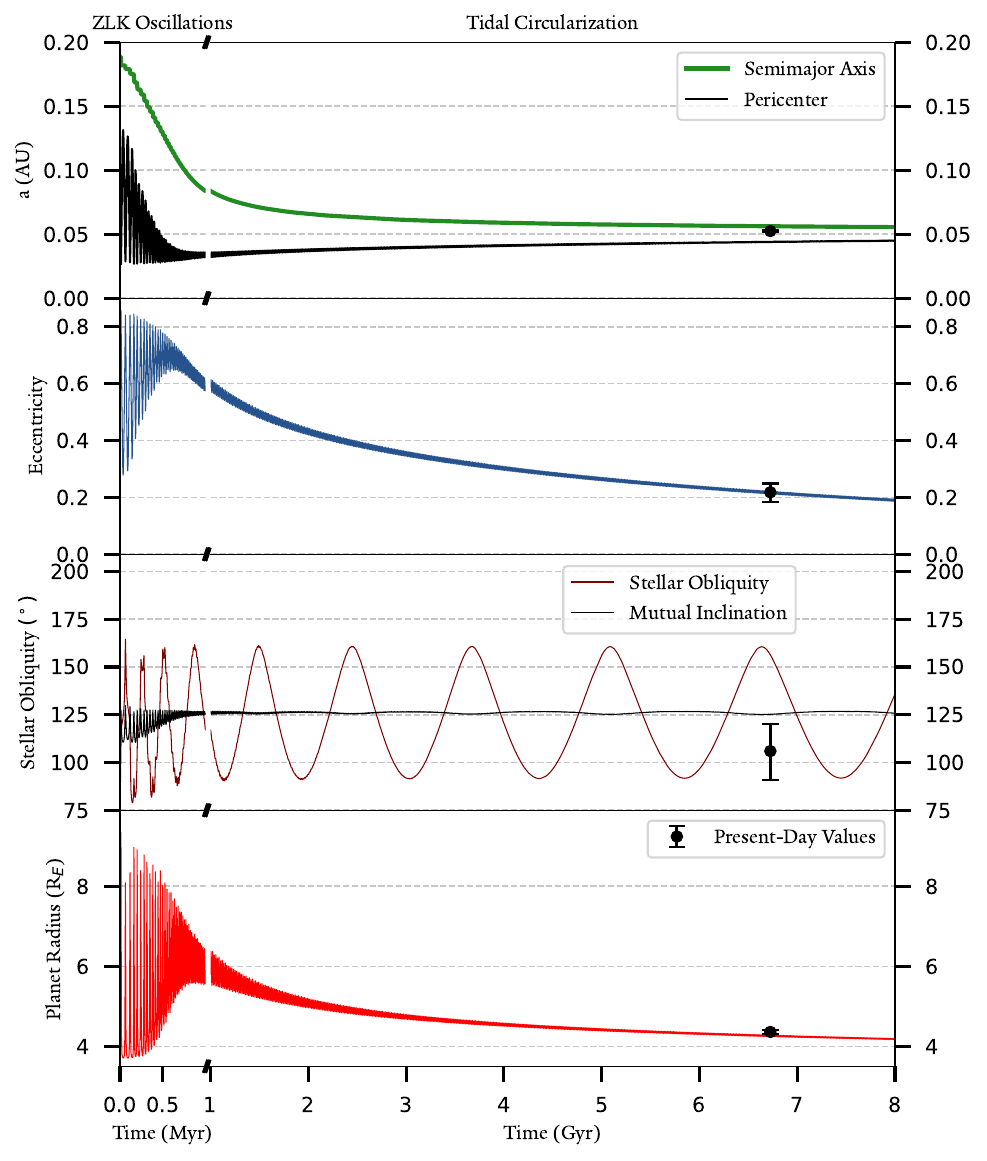}
    \caption{Fiducial ZLK migration of HAT-P-11 b. From top to bottom, HAT-P-11 b's semimajor axis, eccentricity, spin-orbit misalignment and radius are shown with the colored lines. The black dots represent HAT-P-11 b's measured orbital parameters, with the error bar indicating the $1\sigma$ confidence interval. We posit that HAT-P-11 b is still in the tidal circularization stage after the suppression of ZLK oscillations. There is excellent agreement between our simulation results and observed values.}
    \label{fig:zlk}
\end{figure*}

Figure \ref{fig:zlk} shows one of our simulations in detail. Table \ref{tab:sim_params} lists the initial orbital parameters of the simulation, as well as the final values of of interest. During the first phase of the simulation, ZLK oscillations are active and the planet experiences the characteristic coupled eccentricity and inclination oscillations. At each high-eccentricity epoch, tides act on the planet to lower its semimajor axis. The radius evolution of the planet is also plotted, and we see that during each high-eccentricity epoch the planet inflates to over twice its present-day radius. Note that the minimum pericenter distance (${\sim}0.02 \ \mathrm{AU}$) is well outside twice of the tidal disruption radius ($r_t = R_p (M_*/M_p)^{1/3}{\sim}0.008 \ \mathrm{AU}$), so we do not include mass loss due to tidal stripping in our model, as the effects are weak \citep{Guillochon11}. Later, the ZLK oscillations are damped during a high eccentricity epoch. Tides continue to act to circularize the planet's orbit. In the meantime, the orbit normal of HAT-P-11 b precesses about the invariant plane of the system (essentially HAT-P-11 c's orbit normal), as expected, resulting in periodic spin-orbit misalignment oscillations. As expected, the stellar spin axis and the orbit of HAT-P-11 c do not change significantly over the course of the integration.

We comment again on the interpretation of timescales, given our choice of $\tau$ in order to speed up computations. Please refer to Section \ref{sec:zlk_setup} for justification regarding the choice of an enhanced dissipation rate. The enhanced dissipation rate affects the phase when tidal dissipation dominates, and thus we scale the timescales on the right-hand side of the panel to the realistic values accordingly. Note that with a more realistic tidal time-lag, we would expect to see more spin-orbit misalignment cycles on the right-hand side of the plot when tides dominate (since the orbital precession frequency does not depend on tidal dissipation and shouldn't be scaled). In addition, we would expect to have more ZLK cycles on the left-hand side, since each cycle leads to a weaker semi-major axis decay. However, the qualitative behavior of the evolution is the same. 

The age of the HAT-P-11 system is not well-constrained: \cite{2018AJ_Yee_hatp11_RVS} propose an age of $6.5^{+5.9}_{-4.1}$ Gyr, while \cite{morton2016} present an estimate of $2.69^{+2.88}_{-1.24}$ Gyr. Our results are consistent with both estimates of the system age. Note that the timescale of ZLK migration varied by around an order of magnitude, depending on initial conditions.

\begin{deluxetable}{lrrrr}
\tablecaption{Relevant \reboundcodename simulation parameters}
\label{tab:sim_params}
\tablehead{
\colhead{Parameter} & \colhead{Initial b} & \colhead{Final b} & \colhead{Initial c} & \colhead{Final c}} 
\startdata
$\mathrm{R}$ ($\mathrm{R_{Earth}}$) & 3.69 & 4.27 & N/A & N/A\\
$a$ (AU) & 0.19 & 0.056 & 4.1 & 4.1\\
$e$ & 0.48 & 0.22 & 0.65 & 0.65 \\
$\psi$ $(^\circ)$ & 82 & 156 & 35.3 & 35.4\\
\enddata
    \tablecomments{Relevant input parameters and output values of the \texttt{REBOUND} simulation shown in Figure \ref{fig:zlk}. The final values are the values at the time corresponding to the black dot in Figure \ref{fig:zlk}, which is where $e_\mathrm{b}$ is equal to the best fit value. The orbit of HAT-P-11 c does not change significantly over the course of the simulation.}
\end{deluxetable}

The present-day observational constraints on HAT-P-11 b's orbit are also shown on Figure \ref{fig:zlk}. Our simulation is able to match all values very well: we reproduce all observed values within $2\sigma$ other than semimajor axis, which is well-reproduced qualitatively. We propose that HAT-P-11 b is currently in this final stage of evolution and is still in the process of tidal circularization. We conclude that ZLK migration coupled with thermally driven radius evolution is capable of explaining the present-day orbit of HAT-P-11 b remarkably well. Again, we emphasize that we did not perform an exhaustive parameter space analysis of this system. Rather, our aim is to qualitatively show that such a process is capable of reproducing HAT-P-11 b's orbit. 

We briefly comment on the sensitivity of our results to the value of the tidal parameter $\tau$, which is almost entirely unconstrained. Of the simulations that successfully reached HAT-P-11 b's present-day configuration, the mean evolution time is 2.44 Gyr and the 2$\sigma$ range is $\{0.06, 16.96\}$ Gyr. The distribution of evolution times is sculpted by the initial configuration of the system. For instance, configurations that are initialized with a higher mutual inclination tend to reach the present-day state faster, and vice-versa. The center of the distribution is sensitive to the value of the tidal parameter $\tau$, which shifts the distribution linearly. We provide a reasonable range of $\tau$ values that are consistent with our simulations by matching the upper limit of our simulation times with the lower limit of the age estimates provided by \cite{2018AJ_Yee_hatp11_RVS}, and vice versa. This gives $\tau \in \{4.84 \times 10^{-11}, 7.06 \times 10^{-8}\}$ years as the range of $\tau$ able to reconcile our results with the reported age of the system. At HAT-P-11 b's present-day semimajor axis, this corresponds to $Q \in \{1.5 \times 10^4, 2.20 \times 10^7\}$.

\section{Conclusions}
\label{sec:conclusions}
We have proposed a two-step process of planet-planet scattering followed by subsequent ZLK migration to explain the unusual architecture of the HAT-P-11 system. A violent scattering history in which two former planets are ejected from the system is a viable explanation for the high eccentricity and spin-orbit misalignment of HAT-P-11 c. Then, the ZLK mechanism follows naturally with large mutual inclination between HAT-P-11 b and c. We find that traditional ZLK migration with fixed planetary radius cannot reproduce the orbit of HAT-P-11 b, but accounting for thermally-driven radius inflation can. Using \textit{N}-body simulations, we have verified that this scenario is consistent with all observational constraints as well as the estimated age of the system. The HAT-P-11 system is thus an excellent case study of planet-planet scattering and ZLK migration, both of which are common effects believed to sculpt the architectures of many planetary systems.


HAT-P-11 b adds to a growing census of planets on polar orbits around their host stars \citep{albrecht_perponderance} -- the exact significance of this peak in the stellar obliquity distribution is sensitive to detection biases and thus has not been definitively confirmed \citep{siegel_ponderings, Dong23}. ZLK migration triggered by the presence of an external planetary perturber has been shown to be theoretically capable of generating the peak of perpendicular planets \citep{petrovich_warm}, and \cite{vick_misaligned_zlk} demonstrated that if the inner planet starts on an initially misaligned orbit, then perpendicular planets are preferentially produced. They attributed this primordial misalignment to an inclined binary companion torquing the protoplanetary disk \citep{spalding_2014, zanazzi_misalignment, gerbig_2024}. We show that planet-planet scattering can lead to similar ZLK initial conditions for single star systems, and hence we demonstrate that the interplay between planet-planet scattering and ZLK migration can easily produce perpendicular planets as well. While spin-orbit misalignment measurements of close-in planets are relatively common, the same cannot be said of the longer-period giants which would constitute said external perturbers \citep{rice_soles}. The possibility of more spin-orbit misalignment measurements of these longer-period planets in the near future will allow more comprehensive analysis of systems characterized by planet-triggered ZLK migration.

Our study demonstrates the importance of considering the evolution of the physical characteristics of a planet during ZLK migration, in addition to the orbital parameters. In the high-eccentricity epochs of ZLK migration, the planet's envelope can inflate due to tidal heating in the interior. This is highly significant, affecting not only the timescale of ZLK migration but also the equilibrium semimajor axis the planet settles at. This is particularly important in the close companion ZLK cases, since the strength of the ZLK perturbations is relatively much stronger. We show that radius inflation unlocks areas of parameter space which otherwise would be impossible to access through standard ZLK migration with fixed radius. HAT-P-11 is one of the first candidates with two confirmed planets in which ZLK migration is hypothesized to have occurred. A few other systems for which planet-planet ZLK may be possible have begun to emerge \citep{beust_2012, Petrovich_2018, bgagliuffi_2021}, and as more such systems are discovered it will be crucial to consider the evolution of the physical parameters of the planet. During the preparation of this manuscript \cite{yu_dai_2024} performed a similar analysis on the WASP-107 system. Their study used a secular orbital evolution code and also accounted for physical processes such as tidal disruption. This work represents another step forward into more nuanced consideration of coupled structure and orbital evolution.

While our study highlights the importance of considering physical evolution of the planet, our prescription is a first-order approximation. Given the significant uncertainties in quantities such as internal composition and tidal quality factor present in all exoplanetary systems we have not explored more sophisticated models, and assert that our first-order approximation provides sufficient qualitative insight. However, future work could more deeply consider avenues such as 1) a more sophisticated tidal model -- calculating tidal heating and orbital evolution based on dynamical tides, which are more accurate for high eccentricity orbits, 2) coupling tidal heating and radius inflation during orbital evolution (e.g., implementing a lag time such that the radius does not respond instantaneously to tidal forcing, and considering the change in planetary structure when calculating tidal effects \citep{ogilvie_2009}) and 3) considering the effects of atmospheric mass loss \citep{vissapragada_2022}.

Finally, we note that while it may be tempting to place constraints on quantities such as the tidal quality factor with this work, this must be done with caution. The final state of HAT-P-11 b depends on the strength of tidal forcing compared to the magnitude of the precession induced by the ZLK effect. In practice, this means that it depends on a complex interplay of the initial semimajor axis, tidal quality factor, tidal love number, initial mutual inclination, and internal composition, none of which are well-constrained. To explore such a large parameter space would be a herculean task, and we again stress that this study does not claim to have done so. Rather, our work should be viewed more qualitatively as a proof-of-concept that thermally-driven radius inflation is highly significant towards the final products of planet-planet ZLK migration.

\software{
          \texttt{numpy} \citep{numpy1, numpy2},
          \reboundcodename \citep{rebound_2012_main}, \reboundxcodename \citep{tamayo2020reboundx}, 
          Jupyter (\url{https://jupyter.org/}).
          }

\acknowledgments
We thank the anonymous referee for comments which greatly improved the manuscript. We thank Doug Lin, Yasmeen Asali, Daniel Fabrycky, Konstantin Gerbig, Evgeni Grishen, Yurou (Nina) Liu, Rosemary Mardling, Michael Poon, Chloe Neufeld, and Yubo Su for insightful discussions. GL is grateful for the support by NASA 80NSSC20K0641 and 80NSSC20K0522. This work has benefited from use of the \texttt{Grace} computing cluster at the Yale Center for Research Computing (YCRC). Some plots in this work have made use of the colormaps in the \texttt{CMasher} package \citep{vanderveldencmasher}.

\section*{Data Availability}
The \texttt{REBOUND} C code used to generate all simulations in this work can be found at \texttt{https://github.com/tigerchenlu98/reboundx/}
\texttt{tree/hatp11\_dynamics}.

\bibliographystyle{aasjournal}
\bibliography{refs.bib}

\appendix
\section{Self-Consistent Spin, Tidal and Dynamical Equations of Motion}
\label{appendix:a}

The ZLK migration results in Section \ref{sec:zlk} account for spin and tidal evolution in the equilibrium tide framework. The equations of motion used are reproduced from \cite{eggleton1998equilibrium}, but see also \cite{alexander1973weak,hut_binary,mardling2002calculating} for a deeper review of equilibrium tide theory. Also see \cite{Lu_2023} for details on the specific implementation.

In addition to point-particle gravity, we consider the acceleration due to the quadrupole distortion of body~1, which accounts for both spin distortion and tidal perturbation from another body~2:

\begin{equation}
    \begin{split}
        \textbf{f}_{\text{QD}}^{(1, 2)} =  r_1^5 k_{L,1}\left(1 + \frac{m_2}{m_1}\right) \cdot \bigg[& \frac{5(\mathbf{\Omega}_1 \cdot \mathbf{d})^2 \mathbf{d}}{2d^7} - \frac{\Omega_1^2 \mathbf{d}}{2d^5} - \frac{(\mathbf{\Omega}_1 \cdot \mathbf{d}) \mathbf{\Omega}_1}{d^5} - \frac{6 G m_2 \mathbf{d}}{d^8}\bigg],
    \end{split}
\end{equation}
with an analogous expression for the quadrupole distortion of body~2 due to body~1. We also consider the acceleration due to the tidal damping of body~1:

\begin{equation}
    \begin{split}
        \textbf{f}_{\text{TF}}^{(1, 2)} = & -\frac{9 \sigma_1 k_{L,1}^2 r_1^{10}}{2 d^{10}} \left(m_2 + \frac{m_2^2}{m_1}\right) \cdot \left[3 \mathbf{d} (\mathbf{d} \cdot \dot{\mathbf{d}}) + (\mathbf{d} \times \dot{\mathbf{d}} - \mathbf{\Omega}_1 d^2) \times \mathbf{d}\right],
    \end{split}
\end{equation}
again, with an analogous expression for the tidal damping of body~2. 

\section{Orbital Parameter Definitions}
\label{appendix:b}
In Equation \eqref{eq:l_averaged} several terms relating to the precise orientation of the orbit are mentioned but not defined. These are provided here. First, we define the spin vector $\mathbf{\Omega}$, which parameterizes both the direction and magnitude of the planet's rotation. Next, we define two vectors $\mathbf{e}, \mathbf{h}$. These are the Runge-Lenz vector (pointing in the direction of periastron, with magnitude equal to the eccentricity) and the orbital angular momentum vector. We define a third vector $\mathbf{q} \equiv \mathbf{e} \times \mathbf{h}$. We can thus define the following projections of the spin axis:

\begin{equation}
    \Omega_e = \mathbf{\Omega} \cdot \mathbf{\hat{e}}, \Omega_h = \mathbf{\Omega} \cdot \mathbf{\hat{h}}, \Omega_q = \mathbf{\Omega} \cdot \mathbf{\hat{q}}
\end{equation}
The functions of eccentricity $h_i(e)$ are also given:

\begin{equation}
    h_1(e) = \frac{1 + (3/2)e^2 + (1/8)e^4}{(1-e^2)^{9/2}}
\end{equation}

\begin{equation}
    h_2(e) = \frac{1 + (9/2)e^2 + (5/8)e^4}{(1-e^2)^{9/2}}
\end{equation}

\begin{equation}
    h_3(e) = \frac{1 + 3e^2 + (3/8)e^4}{(1-e^2)^{9/2}}
\end{equation}

\begin{equation}
    h_4(e) = \frac{1 + (15/2)e^2 + (45/8)e^4 + (5/16)e^6}{(1-e^2)^6}
\end{equation}

\begin{equation}
    h_5(e) = \frac{1 + (31/2)e^2 + (255/8)e^4 + (185/16)e^6 + (25/64)e^8}{(1-e^2)^{15/2}}
\end{equation}

\end{document}